\documentclass[pra,aps,twocolumn,superscriptaddress,amsmath,amssymb, showpacs]{revtex4-1}

\usepackage{graphicx}

\usepackage{color}

\newcommand{\ket}[1]{|#1\rangle}

\newcommand{\be}{\begin{equation}}
\newcommand{\ee}{\end{equation}}
\newcommand{\bea}{\begin{eqnarray}}
\newcommand{\eea}{\end{eqnarray}}

\newcommand{\kommentar}[1]{}

\begin{document}

\title{Quantum fidelity of symmetric multipartite states}

\author{A. Neven}
\affiliation{CESAM Research Unit, Institut de Physique Nucl\'eaire,
Atomique et de Spectroscopie, University of Liege, Li\`ege 4000, Belgium}

\author{P. Mathonet}
\affiliation{D\'epartement de Math\'ematique, University of Liege (ULg),
Li\`ege 4000, Belgium}

\author{O. G\"uhne}
\affiliation{Naturwissenschaftlich-Technische Fakult\"at, Universit\"at Siegen,
Walter-Flex-Stra{\ss}e 3, D-57068 Siegen, Germany}

\author{T. Bastin}
\affiliation{CESAM Research Unit, Institut de Physique Nucl\'eaire,
Atomique et de Spectroscopie, University of Liege, Li\`ege 4000, Belgium}

\date{28 November 2016}

\begin{abstract}
For two symmetric quantum states one may be interested
in maximizing the overlap under local operations applied to one of them. The question
arises whether the maximal overlap can be obtained by applying the same local operation
to each party. We show that for two symmetric multiqubit states and local unitary transformations this is the case; the maximal overlap can be reached by applying the
same unitary matrix everywhere.  For local invertible operations (stochastic local operations assisted by classical communication equivalence), however, we present counterexamples,
demonstrating that considering the same operation everywhere is not enough.
\end{abstract}

\pacs{03.65.Ud, 03.67.Mn}

\maketitle

\section{Introduction}

In many applications and implementations of quantum information processing,
one has to compare two different quantum states. In this context, the quantum fidelity~\cite{Uhl76, Joz94}
is a very useful tool to measure the ``closeness'' between two states in the Hilbert space
of a quantum system. For two arbitrary states $\rho_1$ and $\rho_2$, it is defined
as
\begin{equation}
F(\rho_1,\rho_2) = \left(\mathrm{Tr} \sqrt{\sqrt{\rho_1} \rho_2 \sqrt{\rho_1}} \right)^2.
\end{equation}
For any pair of pure states $|\psi\rangle$ and $|\phi\rangle$, the quantum
fidelity reduces to their (squared) overlap,
$F(\psi,\phi) = |\langle \psi | \phi \rangle|^2$. Although the fidelity
does not define a metric on the state space, it is the core ingredient for several
of them, like for instance the Bures distance $d_B(\rho_1, \rho_2) =
[2-2\sqrt{F(\rho_1,\rho_2)}]^{1/2}$~\cite{Bur69, Hub9293}. The fidelity is also widely used to define
various entanglement monotones. The Bures distance $d_B$ is itself such an
example~\cite{Som03}. For multipartite pure states, the geometric measure of
entanglement $E_G(\psi) = 1 - F_{|\psi\rangle, \mathcal{S}}$ with
\begin{equation}
\label{FS}
F_{|\psi\rangle, \mathcal{S}} \equiv \sup_{|\phi\rangle \, \mathrm{sep}} F(\psi,\phi)
\end{equation}
is another example that exploits the maximal fidelity between the state
$|\psi\rangle$ to characterize and the set $\mathcal{S}$ of all
fully separable states $\ket{\phi} = \ket{a}\otimes \ket{b} \otimes\ket{c}
\otimes \cdots $~\cite{Wei03}.

The question arises as to how the supremum in Eq.~(\ref{FS}) can be computed. In general, it is known to be an NP-hard task~\cite{Hua14}. For multipartite states
$|\psi_S\rangle$ that are symmetric with respect to the permutations of the
parties it has been shown~\cite{Hub09} that this supremum is realized among
the symmetric separable states $\ket{\phi} = \ket{a}\otimes \ket{a} \otimes\ket{a} \otimes \cdots $ only,
\begin{equation}
\label{FSS}
F_{|\psi_S\rangle, \mathcal{S}}
=
F^S_{|\psi_S\rangle, \mathcal{S}} \equiv \sup_{\mathrm{symmetric} \, |\phi\rangle \, \mathrm{sep}} F(\psi_S,\phi).
\end{equation}
In fact it can even be proven that for three or more particles the state
maximizing the overlap in the definition of $F_{|\psi_S\rangle, \mathcal{S}}$ is necessarily
symmetric \cite{Hub09}. This nice property considerably simplifies the calculation of the
geometric measure of entanglement for symmetric states.

The maximization of the fidelity over sets other than the separable states
$\mathcal{S}$ has proven to be very useful for discrimination strategies of
inequivalent classes of multipartite entangled states with witnesses~\cite{acin00}
or other methods~\cite{Nie10}. In this case, the maximization is typically
to be performed on sets of states equivalent through either local unitary
operations (LU) or stochastic local operations assisted by classical
communication (SLOCC). One needs to evaluate the maximal fidelity
\begin{equation}
\label{FC}
F_{|\psi\rangle, \mathcal{C}} \equiv \sup_{|\phi\rangle \in \mathcal{C}} F(\psi,\phi)
\end{equation}
with $\mathcal{C}$ any considered LU or SLOCC class of states. Mathematically, these classes are defined as follows:
The LU equivalence class of a pure state $\ket{\chi}$ is given by all states
of the form $\ket{\phi}= U_1 \otimes U_2 \otimes \dots \otimes U_N \ket{\chi}$,
where the $U_k$ are unitary matrices acting on the $k$-th party. The SLOCC
equivalence class of $|\chi\rangle$ is given by normalized states of the form $\ket{\phi} \sim A_1 \otimes A_2 \otimes \dots \otimes A_N \ket{\chi}$, where the $A_k$ are invertible matrices \cite{dur99}. The LU and SLOCC equivalence classes of states never coincide, except for the fully separable states that are all both LU and SLOCC equivalent.

In Eq.~(\ref{FC}), if $\mathcal{C}$ contains symmetric states, the question naturally arises whether
the simplification given by Eq.~(\ref{FSS}) for the particular case of the SLOCC (and LU) class $\mathcal{S}$ of separable states generalizes similarly. In other words,
do we have for any symmetric state $|\psi_S\rangle$ and any LU or SLOCC classes $\mathcal{C}$ containing symmetric states
\begin{equation}
\label{FCC}
F_{|\psi_S\rangle, \mathcal{C}} \stackrel{?}{=}
F^S_{|\psi_S\rangle, \mathcal{C}} \equiv
\sup_{\mathrm{symmetric} \, |\phi\rangle \in \mathcal{C}} F(\psi_S,\phi)?
\end{equation}

This paper provides answers to this question for multiqubit systems.
First, in the case of LU classes, the answer is positive and this is
formally proven in Sec.~\ref{LUclassSec}. Second, for the case of SLOCC classes, the answer
is surprisingly negative and spectacular violations of Eq.~(\ref{FCC})
will be given in Sec.~\ref{SLOCCclassSec}. In Sec.~\ref{ConclusionSection} we summarize and discuss further
open problems.

\section{Case of local unitary transformations}
\label{LUclassSec}

When considering LU equivalence classes $\mathcal{C}$ and multiqubit
systems in Eq.~(\ref{FCC}) and since any two LU-equivalent symmetric states can be transformed into each other with the same local unitary acting on each party~\cite{Cen10,Mig13}, the question can be rephrased as follows: Do we
have, for any $N$-qubit symmetric states $|\psi_S\rangle$ and
$|\phi_S\rangle$,
\begin{align}
\label{supSLU2}
    & \sup_{U_1, \ldots, U_N \in \mathrm{U(2)}} |\langle \psi_S | U_1 \otimes \cdots \otimes U_N | \phi_S \rangle |^2 \nonumber \\
    & \qquad = \sup_{U \in \mathrm{U(2)}} |\langle \psi_S | U^{\otimes N} | \phi_S \rangle |^2,
\end{align}
where $\mathrm{U(2)}$ is the group of unitary matrices of dimension $2 \times 2$? Since only the absolute value of the overlap matters, we can choose the phases of
the $U_k$ as we like. So, it suffices to take matrices with determinant $1$ and
consider
\begin{align}
    & \sup_{U_1, \ldots, U_N \in \mathrm{U(2)}} |\langle \psi_S | U_1 \otimes \cdots \otimes U_N | \phi_S \rangle |^2 \nonumber \\
    & \quad = \sup_{U_1, \ldots, U_N \in \mathrm{SU(2)}} |\langle \psi_S | U_1 \otimes \cdots \otimes U_N | \phi_S \rangle |^2,
\end{align}
with $\mathrm{SU(2)}$ the group of unitary matrices of determinant~1.
The question (\ref{supSLU2}) can thus be rephrased as
\begin{align}
\label{supSLSU2}
    & \sup_{U_1, \ldots, U_N \in \mathrm{SU(2)}} |\langle \psi_S | U_1 \otimes \cdots \otimes U_N | \phi_S \rangle |^2 \nonumber \\
    & \qquad = \sup_{U \in \mathrm{SU(2)}} |\langle \psi_S | U^{\otimes N} |\phi_S \rangle |^2 ?
\end{align}

To tackle this problem, we note that an arbitrary $\mathrm{SU(2)}$ matrix can be written
as
\begin{equation}
    U = \left( \begin{array}{cc} \alpha & -\beta^* \\ \beta & \alpha^* \end{array}\right), \quad \alpha, \beta \in \mathbb{C} : |\alpha|^2 + |\beta|^2 = 1.
\end{equation}
We then define the function
\begin{align}
    & P_{\psi_S, \phi_S} : \mathbb{C}^2 \times \cdots \times \mathbb{C}^2 \rightarrow \mathbb{C} : (q_1, \ldots ,q_N) \rightarrow \nonumber \\
    & \quad  P_{\psi_S, \phi_S}(q_1, \ldots, q_N) = \langle \psi_S | U_1 \otimes \cdots \otimes U_N | \phi_S \rangle,
\end{align}
with
\begin{equation}
    U_i = \left( \begin{array}{cc} \alpha_i & -\beta_i^* \\ \beta_i & \alpha_i^* \end{array}\right) : (\alpha_i, \beta_i) \equiv q_i,
\end{equation}
where $\mathbb{C} \simeq \mathbb{R}^2$ and $\mathbb{C}^2 \simeq \mathbb{R}^4$
are considered as \emph{real} Hilbert spaces. The function $P_{\psi_S, \phi_S}$
is symmetric with respect to the permutations of the variables, because $|\psi_S\rangle$ and $|\phi_S\rangle$ are symmetric states. Furthermore, it is $\mathbb{R}$-multilinear
in the coefficients $q_1, \ldots, q_N$, that is,
\begin{align}
    \forall & \; s, t \in \mathbb{R}, \;\;   i = 1, \dots, N: \nonumber \\
    & \quad P_{\psi_S, \phi_S}(q_1, \ldots, s q_i + t p_i , \ldots, q_N) \\
    & \quad = s P_{\psi_S, \phi_S}(q_1, \ldots, q_N) + t P_{\psi_S, \phi_S}(q_1, \ldots, p_i, \ldots, q_N). \nonumber
\end{align}
The multilinearity is only ensured for real $s$ and $t$ and this is the reason why we have to consider the \emph{real} Hilbert spaces $\mathbb{C}$ and $\mathbb{C}^2$.
Under these conditions, H\"ormander's theorem 4 of Ref.~\cite{Hor54} and its extension to the case of real Hilbert spaces of any dimension~\cite{Hor54, Kel28}
can be applied and we have
\begin{equation}
    \label{Horm}
    \sup_{q_1, \ldots, q_N \in \mathbb{C}^2} \frac{|P_{\psi_S, \phi_S}(q_1, \ldots, q_N)|^2}{\Vert q_1\Vert^2 \cdots \Vert q_N \Vert^2} = \sup_{q \in \mathbb{C}^2} \frac{|P_{\psi_S, \phi_S}(q)|^2}{\Vert q \Vert^{2N}},
\end{equation}
where $P_{\psi_S, \phi_S}(q) \equiv P_{\psi_S, \phi_S}(q, \ldots, q)$. Because of the $\mathbb{R}$-multilinearity of $P_{\psi_S, \phi_S}$,
\begin{equation}
    \frac{|P_{\psi_S, \phi_S}(q_1, \ldots, q_N)|^2}{\Vert q_1\Vert^2 \cdots \Vert q_N\Vert^2} = |P_{\psi_S, \phi_S}(q'_1, \ldots, q'_N)|^2,
\end{equation}
with $q'_i = q_i/\Vert q_i \Vert$ and thus
\begin{align}
    & \sup_{q_1, \ldots, q_N \in \mathbb{C}^2} \frac{|P_{\psi_S, \phi_S}(q_1, \ldots, q_N)|^2}{\Vert q_1\Vert^2 \ldots \Vert q_N\Vert^2} \nonumber \\
    & \quad = \sup_{q_1, \ldots, q_N \in \mathbb{C}^2 : \Vert q_i\Vert^2 = 1} |P_{\psi_S, \phi_S}(q_1, \ldots, q_N)|^2.
\end{align}
Equation~(\ref{Horm}) then yields
\begin{align}
    \label{Horm2}
    & \sup_{q_1, \ldots, q_N \in \mathbb{C}^2 : \Vert q_i\Vert^2 = 1} |P_{\psi_S, \phi_S}(q_1, \ldots, q_N)|^2 \nonumber \\
    & \quad = \sup_{q \in \mathbb{C}^2 : \Vert q\Vert^2 = 1} |P_{\psi_S, \phi_S}(q)|^2,
\end{align}
that is to say,
\begin{align}
    \label{Horm3}
    & \sup_{U_1, \ldots, U_N \in \mathrm{SU(2)}} |\langle \psi_S | U_1 \otimes \cdots \otimes U_N | \phi_S \rangle |^2 \nonumber \\
    & \quad = \sup_{U \in \mathrm{SU(2)}} |\langle \psi_S | U^{\otimes N} | \phi_S \rangle |^2.
\end{align}
So, the answer to question posed in Eq.~(\ref{supSLSU2}) and in Eq.~(\ref{supSLU2}) is clearly positive. This finishes the proof.

\section{Case of SLOCC transformations}
\label{SLOCCclassSec}

Now we turn to the case when SLOCC classes are considered in Eq.~(\ref{FCC}). Since
any two SLOCC-equivalent symmetric states can be transformed into each other with the same invertible local operation (ILO) acting on each party~\cite{Math10, Mig13}, the question addressed here for multiqubit
systems is the following: Do we have, for any $N$-qubit symmetric states $|\psi_S\rangle$ and
$|\phi_S\rangle$,
\begin{align}
\label{supSLOCC}
    & \sup_{A_1, \ldots, A_N \in \mathrm{GL(2)}} \frac{|\langle \psi_S | A_1 \otimes \cdots \otimes A_N | \phi_S \rangle |^2}{\| A_1 \otimes \cdots \otimes A_N \ket{\phi_S}\|^2} \nonumber \\
    & \qquad = \sup_{A \in \mathrm{GL(2)}} \frac{|\langle \psi_S | A^{\otimes N} | \phi_S \rangle |^2}{\| A^{\otimes N} \ket{\phi_S} \|^2},
\end{align}
where $\mathrm{GL(2)}$ is the group of invertible matrices of dimension $2 \times 2$?

Here, the expression to be maximized contains a normalization constant that also depends on the state $|\phi_S\rangle$. Contrary to the LU case, the left-hand side term of Eq.~(\ref{supSLOCC}) cannot be cast in a multilinear form divided by a product of norms like the left-hand side term of Eq.~(\ref{Horm}). H\"ormander's theorem therefore cannot be exploited to tentatively prove Eq.~(\ref{supSLOCC}). Actually, the counterexamples to this equation identified hereafter prove that any attempt in this direction would be doomed to failure.

As the first observation, for three-qubit systems, extensive numerical
simulations showed no violation of Eq.~(\ref{FCC}). Similarly, numerical
simulations for $N$ up to 8 gave indications that Eq.~(\ref{FCC}) seems
also to hold for the classes of states SLOCC-equivalent to the $1$-excitation Dicke states $|D_N^{(1)}\rangle$~\cite{footnote1}, hereafter denoted by the classes $\mathcal{W}_N^{(1)}$. These classes gather both symmetric and nonsymmetric states. When restricted to the symmetric subspace, they merely identify to the $\mathcal{D}_{N-1,1}$ families of symmetric states of Ref.~\cite{Bas09}. This brings us to conjecture that the equality
\begin{equation}
    \label{conj}
    F_{|\psi_S\rangle, \mathcal{W}_N^{(1)}} = F^{S}_{|\psi_S\rangle, \mathcal{W}_N^{(1)}}
\end{equation}
actually holds for any $N$ and any symmetric state $|\psi_S\rangle$.

\subsection{First counterexample}

\begin{figure}[t!]
	\centering
	\includegraphics[width=\columnwidth,bb=0 280 425 730,clip = true]{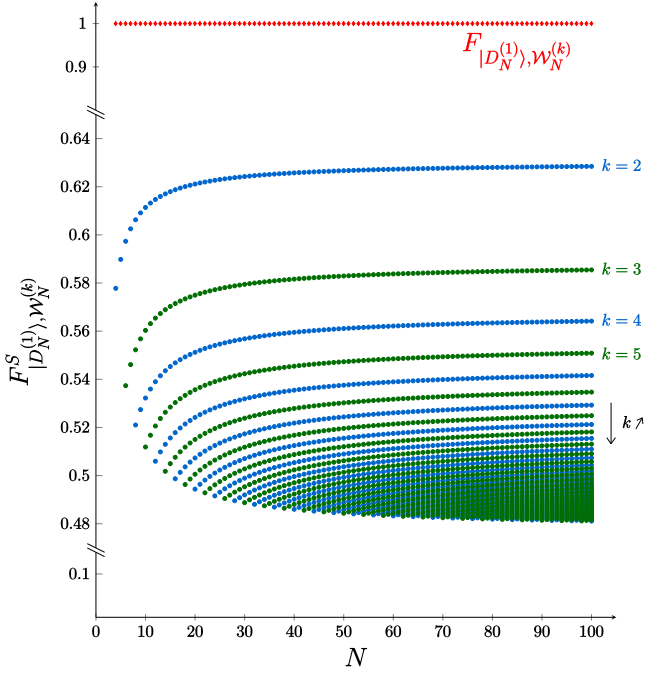}
	\caption{(Color online) Quantum fidelities $F^S_{|D_N^{(1)}\rangle, \mathcal{W}_N^{(k)}}$ (green and blue dotted points) and $F_{|D_N^{(1)}\rangle, \mathcal{W}_N^{(k)}}$ (red diamond points) as a function of $N \geqslant 4$ for each $k = 2, \ldots, \lfloor N/2 \rfloor$. For a given $k$ value, the first plotted quantum fidelity $F^S$ is for $N = 2 k$. \label{FSDN1DNmkk}}
\end{figure}

The generalization of Eq.~(\ref{conj}) to arbitrary SLOCC classes containing symmetric states, however, is not correct. Spectacular violations are obtained when considering the classes of states SLOCC-equivalent to the $k$-excitation Dicke states $|D_N^{(k)}\rangle$~\cite{footnote1}, hereafter denoted by the classes $\mathcal{W}_N^{(k)}$, for $N \geqslant 4$ and $k = 2, \ldots, \lfloor N/2 \rfloor$. All these classes gather both symmetric and nonsymmetric states. When restricted to the symmetric subspace, they identify to the $\mathcal{D}_{N-k,k}$ families of symmetric states of Ref.~\cite{Bas09}. For the values of $N$ and $k$ considered, there are symmetric states $|\psi_S\rangle$ for which
 \begin{equation}
    F_{|\psi_S\rangle, \mathcal{W}_N^{(k)}} > F^{S}_{|\psi_S\rangle, \mathcal{W}_N^{(k)}}.
\end{equation}
To prove this result, we first consider the state
$|\psi_S\rangle = |D_N^{(1)}\rangle \in \mathcal{W}_N^{(1)}$. For all
aforementioned $N$ and $k$, $\mathcal{W}_N^{(1)} \neq \mathcal{W}_N^{(k)}$~\cite{footnote1}
and one gets (see Appendix~\ref{ApA} for a detailed calculation)
\begin{equation}
\label{rSNk}
F^S_{|D_N^{(1)}\rangle, \mathcal{W}_N^{(k)}} = N C_{N}^{k} \mathop{\sup_{x,x' \in [0,1]}}_{y \in [-1,1]} f(x,x',y)
\end{equation}
with
\begin{widetext}
\begin{equation}
    f(x,x',y) =  (1 - x^2)^{N-k-1}(1 - x'^2)^{k-1} \frac{\left(\frac{N-k}{N}\right)^2 x^2 (1-x'^2) + \left(\frac{k}{N}\right)^2 x'^2 (1-x^2) + 2 \frac{k}{N}\frac{N-k}{N} y x x' \sqrt{1-x^2}\sqrt{1-x'^2}}{\sum_{j=0}^{k} C_{k}^j C_{N-k}^j \left[ x^2 x'^2 + (1-x^2)(1-x'^2) + 2 y x x' \sqrt{(1-x^2)(1-x'^2)})\right]^{j}}.
\end{equation}
\end{widetext}
Equation~(\ref{rSNk}) cannot be reduced analytically. It can, however, be straightforwardly evaluated numerically. We illustrate it in Fig.~\ref{FSDN1DNmkk} for
$N = 4, \dots, 100$ and $k = 2, \dots, \lfloor N/2 \rfloor$. The figure
clearly shows that $F^S_{|D_N^{(1)}\rangle, \mathcal{W}_N^{(k)}}$ remains
significantly below one with an asymptotic behavior as $N$ tends to infinity
for fixed $k$. For fixed $N$, the considered fidelities decrease with
increasing $k$. The largest ones are obtained for $k = 2$ with an
asymptotic value for large $N$ of $\sim0.63$.

By contrast, surprisingly we have
\begin{equation}
    \label{maxrNk}
    F_{|D_N^{(1)}\rangle, \mathcal{W}_N^{(k)}} = 1, \quad \forall N \geqslant 4, k = 2, \ldots, \lfloor N/2 \rfloor,
\end{equation}
which means that the Dicke state $|D_N^{(1)}\rangle$ can be approached as closely as desired by non-symmetric SLOCC inequivalent $\mathcal{W}_N^{(k)}$ states. We thus
clearly have
\begin{align}
    F_{|D_N^{(1)}\rangle, \mathcal{W}_N^{(k)}} & > F^{S}_{|D_N^{(1)}\rangle, \mathcal{W}_N^{(k)}}, \nonumber \\
    & \quad \forall N \geqslant 4, k = 2, \ldots, \lfloor N/2 \rfloor,
\end{align}
and this is a neat violation of Eq.~(\ref{FCC}).

To prove Eq.~(\ref{maxrNk}), we define for any $\epsilon \neq 0$ the non-symmetric $\mathcal{W}_N^{(k)}$ state
\begin{equation}
    \label{psiNk}
    |\psi_N^{(k)}({\epsilon}) \rangle \equiv \frac{g_{N,k}({\epsilon}) |D_N^{(k)}\rangle}{\|g_{N,k}({\epsilon}) \ket{D_N^{(k)}} \|},
\end{equation}
with $g_{N,k}(\epsilon)$ the $N$-qubit non-symmetric local operation
\begin{equation}
    g_{N,k}({\epsilon}) = A({\epsilon})^{\otimes N-1} \otimes B_{N,k}({\epsilon}),
\end{equation}
where, in the computational basis $\{|0\rangle, |1\rangle\}$,
\begin{equation}
    A({\epsilon}) = \left( \begin{array}{cc} 1 & 1 \\ -\epsilon & 0 \end{array} \right), \quad B_{N,k}({\epsilon}) = \left( \begin{array}{cc} 1 & 1-\frac{N}{k} \\ \epsilon & \frac{N-2}{N-1}\left(1 - \frac{N}{k} \right) \epsilon \end{array} \right).
\end{equation}
The state $|\psi_N^{(k)}({\epsilon}) \rangle$ belongs to the $\mathcal{W}_N^{(k)}$ SLOCC class since for any $\epsilon \neq 0$ $g_{N,k}({\epsilon})$ is an invertible local operation:
\begin{equation}
    \det A({\epsilon}) = \epsilon \neq 0, \quad \det B_{N,k}({\epsilon}) = \frac{1}{k}\frac{N-k}{N-1} \epsilon \neq 0.
\end{equation}
The state $|\psi_N^{(k)}({\epsilon}) \rangle$ is also non-symmetric since a non-symmetric ILO acting on $|D_N^{(k)}\rangle$ always yields a non-symmetric state for $k \geqslant 2$~\cite{Math10}. A detailed calculation then gives, for any $\epsilon \neq 0$~\cite{footnote2}, (see Appendix~\ref{ApB})
\begin{equation}
    \label{psiNk2}
    |\psi_N^{(k)}(\epsilon)\rangle = \mathcal{N}_{\epsilon} ( |D_N^{(1)}\rangle + |\psi_{\epsilon}\rangle ),
\end{equation}
where $\mathcal{N}_{\epsilon} = 1/\sqrt{1 + \|\psi_{\epsilon}\|^2}$ and
\begin{equation}
\label{psieps}
|\psi_{\epsilon}\rangle = \sum_{j=1}^{N-k} (-\epsilon)^j a_j (b_{j} |D_{N-1}^{(j+1)} \rangle \otimes |0\rangle + c_{j} |D_{N-1}^{(j)}\rangle \otimes |1\rangle),
\end{equation}
with
\begin{align}
    a_j & = \frac{C_{N-j-1}^{k-1}}{C_{N-2}^{k-1}} \sqrt{\frac{C_{N-1}^j}{N}}, \nonumber \\
    b_j & = (N-j-k)\sqrt{\frac{j+1}{N-j-1}}, \label{ajbjcj}\\
    c_j & = \frac{(j-k)-N(j-1)}{N-1}. \nonumber
\end{align}

We have $\lim_{\epsilon \rightarrow 0} |\psi_\epsilon\rangle = 0$ and thus
\begin{equation}
    \lim_{\epsilon \rightarrow 0} |\psi_N^{(k)}(\epsilon)\rangle = |D_N^{(1)}\rangle.
\end{equation}
This implies that
\begin{equation}
    \lim_{\epsilon \rightarrow 0} |\langle D_N^{(1)} | \psi_N^{(k)}(\epsilon)\rangle|^2 = 1^-,
\end{equation}
and this proves Eq.~(\ref{maxrNk}).

These results show that for any $N \geqslant 4$ and $k = 2, \ldots, N-2$, a
non-symmetric ILO of the form $A^{\otimes N-1} \otimes B$ can transform the
Dicke state $|D_N^{(k)}\rangle$ into a non-symmetric state located as closely
as desired to the Dicke state $|D_N^{(1)}\rangle$, even though
$|D_N^{(k)}\rangle$ and $|D_N^{(1)}\rangle$ are SLOCC inequivalent.
This result cannot be achieved when only symmetric SLOCC operations
are considered.

As an aside and out of curiosity, it is interesting to study the
inverse operation $g_{N,k}({\epsilon})^{-1}$ acting on the
$|D_N^{(1)}\rangle$ state. While one obviously has
\begin{equation}
\frac{g_{N,k}({\epsilon})^{-1} |\psi_N^{(k)}(\epsilon)\rangle}{\|g_{N,k}({\epsilon})^{-1} \ket{\psi_N^{(k)}(\epsilon)} \|} = |D_N^{(k)}\rangle,
\end{equation}
the state
\begin{equation}
|\psi_{N,k}^{(1)}(\epsilon)\rangle \equiv
\frac{g_{N,k}({\epsilon})^{-1} |D_N^{(1)}\rangle}
{\|g_{N,k}({\epsilon})^{-1} \ket{D_N^{(1)}} \|}
\end{equation}
reads, up to a normalization constant,
\begin{align}
    & |\psi_{N,k}^{(1)}(\epsilon)\rangle = \\
    & \quad (N-2) k |1,\ldots,1\rangle + (N-3)(N-k) |1,\ldots,1,0\rangle \nonumber \\
    & \quad - |D_{N-1}^{(N-2)}\rangle \otimes \left[\frac{(N-2)(N-k)}{\sqrt{N-1}} |0\rangle + \sqrt{N-1} k |1\rangle\right], \nonumber
\end{align}
This state is totally independent of $\epsilon$ and differs significantly from $|D_N^{(k)}\rangle$. It has only components onto states with at least $N-2$
excitations. Therefore, except for $N=4 and k=2$, $|\psi_{N,k}^{(1)}(\epsilon)\rangle$
has no overlap with $|D_N^{(k)}\rangle$. For $N=4 and k=2$, the fidelity with the
state $|D_N^{(k)}\rangle$ amounts only to $1/14$.

\subsection{More general counterexamples}

\begin{table}[t!!]
    \renewcommand{\arraystretch}{1.3}
    $$
    \begin{array}{|c|c|c||c|c|c|}
    \hline
        N & k & F^S_{|D_N^{(k)}\rangle, \mathcal{W}_N^{(1)}} & N & k & F^S_{|D_N^{(k)}\rangle, \mathcal{W}_N^{(1)}} \\ [1 ex]
        \hline \hline
        4 & 2 & 0.5   & 7 & 2 & 0.457 \\
        5 & 2 & 0.477 &   & 3 & 0.383 \\
        6 & 2 & 0.465 & 8 & 2 & 0.451 \\
          & 3 & 0.4   &   & 3 & 0.372 \\
          &   &       &   & 4 & 0.344\\
          \hline
    \end{array}
    $$
    \renewcommand{\arraystretch}{1}
\caption{Exemplification of some fidelities $F^S_{|D_N^{(k)}\rangle, \mathcal{W}_N^{(1)}}$.}
\label{TabI}
\end{table}

Equation~(\ref{maxrNk}) can even be generalized. Any
state $|\psi_{N,1}\rangle$ of the $\mathcal{W}_N^{(1)}$ SLOCC class satisfies
\begin{equation}
    \label{genN1}
    F_{|\psi_{N,1}\rangle, \mathcal{W}_N^{(k)}} = 1, \quad \forall N \geqslant 4, k = 2, \ldots, \lfloor N/2 \rfloor.
\end{equation}
Indeed, for such states an ILO $h$ can be found such that
\begin{equation}
    |\psi_{N,1}\rangle = h |D_N^{(1)}\rangle.
\end{equation}
We then define for any $\epsilon \neq 0$ the $\mathcal{W}_N^{(k)}$ states
\begin{equation}
    |{\psi'}_N^{(k)}(\epsilon)\rangle = \frac{h |\psi_N^{(k)}(\epsilon)\rangle}{\| h \psi_N^{(k)}(\epsilon) \|},
\end{equation}
with $|\psi_N^{(k)}(\epsilon)\rangle$ as defined by Eq.~(\ref{psiNk2}). We get trivially, up to a normalization constant,
\begin{equation}
    |{\psi'}_N^{(k)}(\epsilon)\rangle = |\psi_{N,1}\rangle + h |\psi_{\epsilon}\rangle,
\end{equation}
and thus
\begin{equation}
    \lim_{\epsilon \rightarrow 0} |{\psi'}_N^{(k)}(\epsilon)\rangle = |\psi_{N,1}\rangle.
\end{equation}
It follows that
\begin{equation}
    \lim_{\epsilon \rightarrow 0} |\langle \psi_{N,1}|{\psi'}_N^{(k)}(\epsilon)\rangle|^2 = 1^-,
\end{equation}
and this implies Eq.~(\ref{genN1}).

This result shows that all states of the $\mathcal{W}_N^{(1)}$ SLOCC
class can be approached as closely as desired by non-symmetric states of
any of the $\mathcal{W}_N^{(k)}$ SLOCC classes, for
$N \geqslant 4 and k = 2, \ldots, \lfloor N/2 \rfloor$. Topologically,
this means that the $\mathcal{W}_N^{(1)}$ SLOCC class of states lies
at the boundary of the non-symmetric side of any of the $\mathcal{W}_N^{(k)}$ classes.
This result sheds an additional light on the general
topology of the SLOCC classes of multiqubit systems, whose restriction on the only symmetric subspace was
established in Ref.~\cite{Bas15}.

The converse of the previous result is by far not true: The states of the $\mathcal{W}_N^{(k)}$ SLOCC classes cannot be approached as closely as desired by $\mathcal{W}_N^{(1)}$ states, even if they are non-symmetric. As an example, a
detailed calculation yields (see Appendix~\ref{ApA})
\begin{equation}
    F^S_{|D_N^{(k)}\rangle, \mathcal{W}_N^{(1)}}=  C_N^k \tilde{k}_r^{k-1} \left[ 1-\tilde{k}_r \right]^{N-k-1} \left[\tilde{k} + (1-2\tilde{k}) \tilde{k}_r \right]
\end{equation}
with $\tilde{k} = k/N$ and $\tilde{k}_r = \tilde{k} - \sqrt{\tilde{k}(1-\tilde{k})/(N-1)}$ and extensive numerical simulations [see our conjecture in Eq.~(\ref{conj})]
showed that this should also correspond to $F_{|D_N^{(k)}\rangle, \mathcal{W}_N^{(1)}}$. We exemplify some values of $F^S_{|D_N^{(k)}\rangle, \mathcal{W}_N^{(1)}}$
in Table~\ref{TabI}. It is noteworthy to mention that these fidelities
decrease with increasing $N$ and $k$. For fixed $k$, we have
\begin{equation}
    \label{limitN}
    \lim_{N \rightarrow \infty} F^S_{|D_N^{(k)}\rangle, \mathcal{W}_N^{(1)}} = \frac{1}{k!} e^{\sqrt{k} - k} (k - \sqrt{k})^k \frac{2 \sqrt{k} - 1}{\sqrt{k} - 1}.
\end{equation}
Explicitly, for $k = 2, \ldots, 8$, this limit reads $0.422$, $0.322$, $0.271$, $0.238$, $0.215$, $0.197$, and $0.183$, respectively. All these results clearly show that while the $|D_N^{(1)}\rangle$ state can be approached as closely as desired by non-symmetric states that are SLOCC equivalent to any of the $|D_N^{(k)}\rangle$ states ($k = 2, \ldots, \lfloor N/2 \rfloor$), none of these latter can be closely approached by any state SLOCC equivalent to the Dicke state $|D_N^{(1)}\rangle$.

\section{Conclusion}
\label{ConclusionSection}
In this paper we have analyzed the maximization
of the quantum fidelity between symmetric multiqubit states and
sets of LU- or SLOCC-equivalent states that contain symmetric
states. We have shown that the open question in Eq.~(\ref{FCC}) admits
a positive answer when LU classes of qubit states are considered,
while the answer is negative when turning to SLOCC classes of states.

In the case of LU sets, the positive answer simplifies considerably the
calculation of the desired maximal overlap. For SLOCC classes of
states, we have shown significant violations of Eq.~(\ref{FCC})
where for some states $|\psi_S\rangle$ and classes $\mathcal{C}$
the fidelity $F_{|\psi_S\rangle, \mathcal{C}}$ takes the maximal
possible value 1, while the symmetric restriction
$F^S_{|\psi_S\rangle, \mathcal{C}}$ has only significantly
much lower values. This is in particular the case when
considering any states $|\psi_S\rangle$ of the
$\mathcal{W}_N^{(1)}$ SLOCC class in combination with
any of the $\mathcal{W}_N^{(k)}$ SLOCC classes,
for $N \geqslant 4$ and $k = 2, \ldots, \lfloor N/2 \rfloor$.
Finally, extensive numerical simulations have also lead
us to conjecture that Eq.~(\ref{FCC}) seems to be
correct when the considered SLOCC class $\mathcal{C}$
identifies to $\mathcal{W}_N^{(1)}$, whatever the
state $|\psi_S\rangle$.

There are several directions in which our work can be generalized.
First, concerning LU equivalence classes, it would be highly desirable
to prove our result also for higher-dimensional systems and not only for qubits.
In our proof, we made use of the simple parametrization of $\mathrm{SU(2)}$ matrices, which
is not so simple in higher dimensional systems. Second, for SLOCC equivalence classes
it would be very useful to find out under which additional conditions the optimization
over symmetric states is enough. Based on numerical evidence  we identified some
examples where this seems to be the case, but so far no clear understanding has been
reached. {From} a more general perspective, our work presents examples where symmetries
can help to solve optimization problems related to the numerical range
\cite{gawron10, gawron11}. Understanding further the role of symmetry in such
problems is clearly a challenging task, nevertheless it will have a significant impact
on various problems in quantum information theory.

\acknowledgments
A.N. acknowledges a FRIA grant and the Belgian F.R.S.-FNRS for financial support. O.G. acknowledges financial support from the FQXi Fund (Silicon Valley Community Foundation), the DFG, and the ERC (Consolidator Grant No. 683107/TempoQ). TB acknowledges financial support from the Belgian F.R.S.-FNRS through IISN Grant No. 4.4512.08.

\appendix

\section{}
\label{ApA}
In this appendix we show that the maximal fidelity $F^S_{|D_N^{(k)}\rangle, \mathcal{W}_N^{(k')}}$ for any $k = 1, \ldots, \lfloor N/2 \rfloor$, $k' = 0, \ldots, \lfloor N/2 \rfloor$, and $k \neq k'$ reads
\begin{widetext}
\begin{equation}
\label{rSNkkp}
     F^S_{|D_N^{(k)}\rangle, \mathcal{W}_N^{(k')}} = \frac{C_N^k}{C_N^{k'}} \mathop{\sup_{x,x' \in [0,1]}}_{y \in [-1,1]} \frac{\sum_{j=0}^{k'} \left[(2 - \delta_{j,0}) \sum_{j' = j}^{k'} c_{j'}(x,x') c_{j'-j}(x,x')\right] T_j(y)}{\sum_{j=0}^{k'} C_{k'}^j C_{N-k'}^j \left[ x^2 x'^2 + (1-x^2)(1-x'^2) + 2 y x x' \sqrt{(1-x^2)(1-x'^2)})\right]^{j}},
\end{equation}
\end{widetext}
where $C_m^n \equiv { m \choose n }$ is the binomial coefficient between $m$ and $n$,
with the usual convention $C_m^n = 0$ if $n < 0$ or $n > m$, $\delta$ denotes the Kronecker delta, $T_j(y)$ is the $j$-th degree Chebyshev polynomial of the first kind, and, for $j = 0, \ldots, k'$,
\begin{align}
\label{cjxxp}
c_j(x,x') = & C_k^j C_{N-k}^{k'-j} x^{k-j} \sqrt{1-x^2}^{N-k'-k+j} \nonumber \\
& \quad \times x'^j \sqrt{1-x'^2}^{k'-j}.
\end{align}

In particular, for $k' = 0$, one gets the well known result~\cite{Wei03}
\begin{align}
    F^S_{|D_N^{(k)}\rangle, \mathcal{W}_N^{(0)} \equiv \mathcal{S}} & = C_N^k \sup_{x \in [0,1]} x^{2k} (1-x^2)^{N-k} \nonumber \\
    & = C_N^k \tilde{k}^k \left(1-\tilde{k}\right)^{N-k},
\end{align}
with $\tilde{k} = k/N$ the fractional excitation of the Dicke state $|D_N^{(k)}\rangle$. For $k' = 1$, one gets
\begin{widetext}
\begin{align}
    F^S_{|D_N^{(k)}\rangle, \mathcal{W}_N^{(1)}} & = \frac{C_N^k}{N} \mathop{\sup_{x,x' \in [0,1]}}_{y \in [-1,1]} \frac{c_0(x,x')^2 + c_1(x,x')^2 + 2 c_0(x,x') c_1(x,x') y}{1 + (N-1)\left[x^2 x'^2 + (1-x^2)(1-x'^2) + 2 y x x' \sqrt{(1-x^2)(1-x'^2)}\right]} \nonumber \\
    & = C_N^k \tilde{k}_r^{k-1} \left( 1-\tilde{k}_r \right)^{N-k-1} \left[\tilde{k} + (1-2\tilde{k}) \tilde{k}_r \right],
\end{align}
\end{widetext}
with
\begin{equation}
\tilde{k}_r = \tilde{k} - \sqrt{\tilde{k}\frac{1-\tilde{k}}{N-1}}.
\end{equation}
For $k' > 1$, Eq.~(\ref{rSNkkp}) cannot be reduced analytically and it must be evaluated numerically. We noticed that for all tested cases the supremum was systematically obtained for $y = -1$.

To prove Eq.~(\ref{rSNkkp}), we first observe that in the computational basis the Dicke state $|D_N^{(k)}\rangle$ merely reads $1/\sqrt{C_N^k} \sum |0,\ldots,0,1,\ldots,1\rangle$, where the sum runs over all multiqubit states with any $k$ qubits in the $|1\rangle$ state and the remaining $N-k$ qubits in the $|0\rangle$ state. We then note that any symmetric state of the $\mathcal{W}_N^{(k')}$ SLOCC class, i.e., states of the $\mathcal{D}_{N-k',k'}$ family, can be written in the form~\cite{Bas09} $\mathcal{N} \sum |\epsilon,\ldots,\epsilon,\epsilon',\ldots,\epsilon'\rangle$, where $\mathcal{N}$ is a normalization constant and the sum runs over all multiqubit states with any $k'$ qubits in an $|\epsilon\rangle$ state and the remaining $N-k'$ qubits in a distinct $|\epsilon'\rangle \neq |\epsilon\rangle$ state. Writing the single qubit states $|\epsilon\rangle$ and $|\epsilon'\rangle$ in the form $\sqrt{1-x^2} |0\rangle + x e^{i \phi} |1\rangle$ and $\sqrt{1-x'^2} |0\rangle + x' e^{i \phi'} |1\rangle$ ($x,x' \in [0,1], \phi,\phi' \in [0,2\pi[$), respectively, yields
$\mathcal{N}^2 = 1/\{C_N^{k'} \sum_{j=0}^{k'}C_{k'}^j C_{N-k'}^j(a(x,x') + b(x,x') \cos \Delta \phi)^j\}$ with $a(x,x') = x^2 x'^2+(1-x^2)(1-x'^2)$, $b(x,x') = 2 x x' \sqrt{1-x^2} \sqrt{1-x'^2}$ and $\Delta \phi = \phi - \phi'$. We then get
\begin{equation}
     F^S_{|D_N^{(k)}\rangle, \mathcal{W}_N^{(k')}} = \mathop{\sup_{x,x' \in [0,1]}}_{ \Delta \phi \in [0,2\pi[} \mathcal{N}^2 C_N^k \left|\sum_{j=0}^{k'} c_j(x,x') e^{i j \Delta \phi}\right|^2,
\end{equation}
with $c_j(x,x')$ as given by Eq.~(\ref{cjxxp}).
Finally, using the identity $\cos(j \Delta \phi) = T_j(\cos \Delta \phi)$ and setting $y = \cos \Delta \phi$ yields Eq.~(\ref{rSNkkp}).

\section{}
\label{ApB}
In this appendix we prove Eq.~(\ref{psiNk2}). We first observe that $A(\epsilon) |0\rangle = |0\rangle - \epsilon |1\rangle$, $A(\epsilon) |1\rangle = |0\rangle$,
and
\begin{align}
\label{BNkeps}
    B_{N,k}(\epsilon) |0\rangle & = |0\rangle + \epsilon |1\rangle, \nonumber \\
    B_{N,k}(\epsilon) |1\rangle & = \left(1-\frac{N}{k}\right)\left(|0\rangle + \epsilon \frac{N-2}{N-1}|1\rangle\right).
\end{align}

We then define the unnormalized Dicke states $|u_N^{(k)}\rangle = \sqrt{C_N^k} |D_N^{(k)}\rangle$,
which satisfy
\begin{equation}
\label{uNkrec}
|u_N^{(k)}\rangle=\left( |u_{N-1}^{(k-1)}\rangle \otimes |1\rangle + |u_{N-1}^{(k)}\rangle \otimes |0\rangle \right).
\end{equation}
It follows that
\begin{align}
g_{N,k}(\epsilon) |u_N^{(k)}\rangle = \left[ A(\epsilon)^{\otimes N-1} |u_{N-1}^{(k-1)}\rangle \otimes B_{N,k}(\epsilon) |1\rangle \right. \nonumber \\
\left. + A(\epsilon)^{\otimes N-1} |u_{N-1}^{(k)}\rangle \otimes B_{N,k}(\epsilon) |0\rangle \right]. \label{eq:decomposition_gD}
\end{align}
Inserting Eq.~(\ref{BNkeps}) in Eq.~(\ref{eq:decomposition_gD}) and observing that for any $N$ and $k$ the symmetric state $A(\epsilon)^{\otimes N} |u_N^{(k)}\rangle$ reads in the Dicke state basis
\begin{equation}
    A(\epsilon)^{\otimes N} |u_N^{(k)}\rangle = C_{N}^{k} \sum_{j=0}^{N-k} (-\epsilon)^j \frac{C_{N-k}^j}{\sqrt{C_N^j}} |D_N^{(j)}\rangle \label{eq:action_of_A}
\end{equation}
yields straightforwardly
\begin{align}
& g_{N,k}(\epsilon) |D_N^{(k)}\rangle = \epsilon \frac{\alpha_{N-1,k}}{\sqrt{\alpha_{N,k}}} \nonumber \\
& \quad \times \sum_{j=0}^{N-k}{ (-\epsilon)^j  a_j \left( b_j |D_{N-1}^{(j+1)}\rangle \otimes |0\rangle + c_j |D_{N-1}^{(j)}\rangle \otimes |1\rangle \right) } \label{gNkeps}
\end{align}
with $a_j$, $b_j$, and $c_j$ as given by Eq.~(\ref{ajbjcj}) and $\alpha_{N,k} = C_N^k/N$. In the sum over $j$ in Eq.~(\ref{gNkeps}), the first term $j = 0$ merely yields the state
$(|D_{N-1}^{(0)}\rangle \otimes |1\rangle + \sqrt{N-1} \: |D_{N-1}^{(1)}\rangle \otimes |0\rangle )/\sqrt{N}$, which is nothing but the Dicke state $|D_N^{(1)}\rangle$ [see Eq.~(\ref{uNkrec})]. The rest of the sum from $j = 1$ to $j = N-k$ is by definition the state $|\psi_\epsilon\rangle$ [Eq.~(\ref{psieps})]. We thus get
\begin{equation}
g_{N,k}(\epsilon) |D_N^{(k)}\rangle = \epsilon \frac{\alpha_{N-1,k}}{\sqrt{\alpha_{N,k}}} \left( |D_N^{(1)}\rangle + |\psi_{\epsilon}\rangle \right),
\end{equation}
from which Eq.~(\ref{psiNk2}) immediately follows for any $\epsilon \neq 0$. For $\epsilon = 0$, $g_{N,k}(\epsilon) |D_N^{(k)}\rangle = 0$ and the normalized state $|\psi_N^{(k)}({\epsilon})\rangle$ is not defined.


\begin{thebibliography}{19}
\bibitem{footnote1}
We recall that the $k$-excitation Dicke states ($k = 0, \ldots, N$) are defined as
$\ket{D_N^{(k)}} = 1/\sqrt{C_N^k} \sum_\pi \ket{0, \dots, 0, 1, \dots, 1}$,
where $C_N^k$ denotes the binomial coefficient ${N \choose k}$, the multiqubit states in the sum contain $k$ qubits in state $|1\rangle$, and $\pi$ denotes all permutations of the qubits leading to different terms in the sum. All $\ket{D_N^{(k)}}$ Dicke states ($k = 0, \ldots, N$) are symmetric and they form an orthonormal basis in the symmetric subspace of the multiqubit system. For $k = 0, \ldots, \lfloor N/2 \rfloor$, where $\lfloor \cdot \rfloor$ denotes the floor function, the Dicke states $|D_N^{(k)}\rangle$ are SLOCC inequivalent between each others~\cite{Bas09}. In contrast, the $|D_N^{(N-k)}\rangle$ and $|D_N^{(k)}\rangle$ states are LU equivalent.
\bibitem{footnote2} Although the $N$-qubit local operation $g_{N,k}(\epsilon)$ remains defined for $\epsilon = 0$, though not invertible in this case, the normalized state $|\psi_N^{(k)}({\epsilon})\rangle$ is in contrast not defined in this case since $g_{N,k}({\epsilon}=0) |D_N^{(k)}\rangle = 0$ (see Appendix~\ref{ApB}).
\end{thebibliography}
\end{document}